\documentclass{article}
\usepackage{amsmath,graphicx,mlspconf}
\usepackage{xcolor}
\usepackage{booktabs}
\usepackage{url}

\toappear{2025 IEEE International Workshop on Machine Learning for Signal Processing, Aug.\ 31-- Sep.\ 3, 2025, Istanbul, Turkey}

\title{Does Language Matter for Early Detection \\ of Parkinson's Disease from Speech?}
\name{
    Peter Plantinga$^{1, 2, 3}$%
    \enspace Briac Cordelle$^{2, 3, 4}$%
    \enspace Dominique Lou\"er$^{1, 2}$%
    \enspace Mirco Ravanaelli$^{2, 3, 4}$%
    \enspace Denise Klein$^{1, 2}$%
    \thanks{Supported by CRBLM (FRQNT/SC); HBHL (McGill University CFREF). Data collection provided by Roozbeh Sattari and Quebec Parkinson Network (QPN). Some compute provided by DRAC (\protect\url{alliancecan.ca}).}
}
\address{
    $^1$ McGill University%
    \qquad $^2$ Centre for Research on Brain, Language, and Music \\
    $^3$ Mila - Quebec Artificial Intelligence Institute%
    \qquad $^4$ Concordia University%
}

\begin{document}

\maketitle

\begin{abstract}
    Using speech samples as a biomarker is a promising avenue for detecting and monitoring the progression of Parkinson's disease (PD), but there is considerable disagreement in the literature about how best to collect and analyze such data. Early research in detecting PD from speech used a sustained vowel phonation (SVP) task, while some recent research has explored recordings of more cognitively demanding tasks. To assess the role of language in PD detection, we tested pretrained models with varying data types and pretraining objectives and found that (1) text-only models match the performance of vocal-feature models, (2) multilingual Whisper outperforms self-supervised models whereas monolingual Whisper does worse, and (3) AudioSet pretraining improves performance on SVP but not spontaneous speech. These findings together highlight the critical role of language for the early detection of Parkinson's disease.
\end{abstract}
\begin{keywords}
speech biomarker, Parkinson's disease, multilingual models, spontaneous speech, linguistic markers
\end{keywords}

\section{Introduction}

Parkinson's disease (PD) affects millions of people worldwide~\cite{marras2018prevalence}. Unfortunately, diagnosing and monitoring this disease can be expensive and time-consuming. However, there is growing evidence that clinicians' efforts could be assisted through the use of speech, as samples are cheap and easy to collect. This line of research has to contend with two major difficulties: the first is developing generalizable solutions from small corpora, and the second is developing human-understandable interpretations of the solution.

While there is some research that suggests that interpretable speech measures may prove accurate enough to serve as a detection system~\cite{shen2024explainable}, methodological errors~\cite{ali2024parkinson} and small evaluation sets have cast doubt on the generalizability of such results~\cite{ramanarayanan2022speech}. Recent research suggests that audio foundation models might provide a more generalizable basis~\cite{favaro2024unveiling}.

A better understanding of the factors that influence foundation models is necessary to support any clinicians who would seek to use these systems. Traditional interpretability techniques fail for detecting PD~\cite{Mancini2024ParkinsonSpeechExplainability}, so we experiment with varying the information content in recordings, the types of encoders, and the language components, to gain insight into what speech properties are attended by audio foundation models.

The contribution of this work is to highlight the role of language from a few different angles. More specifically:

\begin{enumerate}
    \item We isolate the effects of vocal changes from linguistic changes in two ways. In one experiment we compare the accuracy of audio encoders on an SVP task (predominantly vocal cues) against text encoders on a picture description task (predominantly linguistic cues). In another experiment, we compare the accuracy of voice encoders against text encoders on the same picture description task (DPT). In both cases, the text encoders are at least as accurate as the audio and voice encoders.
    \item We compare the accuracy of frozen audio foundation model encoders with simple classifiers and we find that the best-performing encoder is multilingual Whisper~\cite{radford2022whisper}, trained to do transcription, translation, and language identification. We also experiment with self-supervised (SSL) encoders pretrained with either audio events (AudioSet) or solely with speech. Models trained on audio events perform better on the SVP task, while speech models do better on DPT.
    \item We compare the accuracy of multilingual encoders against English-trained encoders. If vocal control is the main source of information for model predictions, then pretraining language should make little difference to prediction accuracy, which is what we find for SSL models. However, if language is an important component of model predictions, then pretraining language should make a significant difference in final accuracy, which is what we find for Whisper.
\end{enumerate}

\noindent Taken together, these three comparisons show that linguistic markers are an important component of identifying signs of Parkinson's disease in speech. This can inform the choice of spoken task to use when collecting data, provide a basis for future language-based investigations of the disease, and inform the choice of model pretraining paradigm when selecting foundation models for this task.

\section{Background}

Current clinical diagnostic procedures for Parkinson's disease are time-intensive and require specialized knowledge. Accuracy of general practitioners and general neurologists using typical techniques such as physical and neurological exams has been estimated at around 76-86\% \cite{diagnostic2014joutsa}. The challenge of diagnosis has led to research in alternative diagnostic methods such as alpha-synuclein seed amplification assay which can aid diagnosis before symptoms begin, with an estimated accuracy around 88\%~\cite{concha2023seed}. A cheaper and faster diagnostic aid from speech would help general neurologists make better diagnoses.

Research in the detection of Parkinson's disease from speech has long focused on the acoustic aspects of speech production over the linguistic aspects. One common task is sustained vowel phonation (SVP), which started as early as 1963~\cite{brown1963organic} and continues through today~\cite{ali2024parkinson}. Another phonation task used for detection is the computation of the Vowel Space Area (VSA) and Vowel Articulation Index (VAI) that measures formant frequency spread for different vowels, which is reduced in PD patients~\cite{skodda2011vowel}. However, there is some research that disputes the effectiveness of vowel-based measures~\cite{rusz2013imprecise}.


There is ongoing research that has found a number of linguistic and paralinguistic changes in PD in addition to vocal changes~\cite{fang2020cognition}. Early stages of PD are associated with reduced prosodic cues, called monopitch~\cite{rusz2024prodromal}. Researchers have also found that Parkinson's patients have word-finding difficulties~\cite{hedman2022word} and that PD affects syntax and word choice~\cite{bocanegra2015syntax} which supports the idea that more cognitively demanding tasks can be better suited for detecting signs of PD. 

Recent attention has started to accrue to the idea that speech foundation models on continuous speech can contribute to a more straightforward and generalizable method for detecting PD~\cite{ali2024parkinson}. Some have found that these models can work even in real-world acoustic conditions with the help of voice activity detection, speech enhancement, and dereverberation models~\cite{la2024exploiting}. Other work has explored strategies for improving generalizability across languages using multiple tasks~\cite{laquatra2025bilingual}, as well as improving the aggregation of dysarthric cues across speech segments with graph neural networks~\cite{sheikh2025graph}.

Unlike these works, our approach is not designed to achieve the highest possible accuracy, as we freeze parameters and add only minimal parameters for prediction. Instead, our goal is to isolate and analyze the influence of linguistic vs. vocal features on model performance.

\section{Methodology}

Our strategy for conducting experiments on linguistic markers in Parkinson's disease is to vary a few key aspects of the model and input space. We explore various:

\begin{enumerate}
    \item \textbf{Tasks} -- sustained vowel phonation (SVP) and a picture description task (DPT).
    \item \textbf{Languages} -- English-only pretrained models and multilingual models.
    \item \textbf{Encoders} -- low-level acoustic features, text encoders, automatic speech recognition (ASR) encoders, and self-supervised (SSL) encoders.
\end{enumerate}

\noindent Taken together, varying these three dimensions reveals patterns indicating the importance of linguistic markers in early detection of PD. Each dimension is explained further below.

\subsection{Speech Tasks}

As mentioned earlier, there is no consensus in the literature about what tasks are most discriminative for the detection of PD. We show experiments with two tasks, a traditional sustained phonation task and a picture description task. The picture description is from the Boston Aphasia Diagnostic Exam~\cite{naeser1978lesion}. Comparing these tasks can help establish whether vocal features or language features are more useful for the detection of PD, and provide a baseline with which to compare different pretrained speech models.

\subsection{Languages}

One important aspect of biomarker design is considerations of universality, and that includes knowing whether pretraining language will make a significant difference on the performance. Speech-based Parkinson's detectors might be highly language-dependent or they might attend vocal or prosodic features that generalize across languages, even in spontaneous speech samples. We test this by experimenting with models pretrained on either multilingual or monolingual data. Particularly Whisper~\cite{radford2022whisper} and XEUS \cite{chen2024xeus} are pretrained on multilingual speech data and Whisper is further trained to complete language-based tasks such as language identification and translation from many languages into English.

\subsection{Audio Encoders}

We conducted experiments using mel filterbank features, OpenSMILE eGeMAPS~\cite{eyben2015geneva} and SpeechBrain VocalFeatures~\cite{speechbrainV1} to show a reasonable baseline performance based on interpretable features.

We also experimented with a large variety of pretrained speech models as encoders. We specifically focused our collection of models to explore the difference between lexically-trained models and models without explicit lexical training. More specifically, we compared automatic speech recognition (ASR) models against self-supervised speech models. We tried state-of-the-art models from both categories.

For the ASR models, we have explored the Whisper, Crisper Whisper~\cite{zusag24crisper}, and Parakeet~\cite{koluguri2024parakeet} models. Crisper Whisper was trained to do transcription and voice activity detection on English and German, and Parakeet was trained to do English transcription only.

For the self-supervised models, we have explored XEUS, WavLM \cite{Chen2021WavLM}, HuBERT \cite{Hsu2021HuBERT}, and wav2vec 2.0 \cite{baevski2020wav2vec}. Out of these, only XEUS is multilingual, the rest are trained on English only. We also experimented with HuBERT and wav2vec 2.0 trained on AudioSet~\cite{gemmeke2017audioset} instead of on speech~\cite{laquatra2024benchmarking}.

We briefly explored NeMo Canary, XLS-R, MMS, and Seamless M4T, but initial experiments did not see good results, perhaps due to some incompatibility in how we structured our experiments. For some models, such as Canary, we suspect that extensive downsampling reduces the performance for our objectives.

For all of our experiments, we froze each model entirely and added only minimal parameters for aggregating features into a prediction. The intention of this approach is to capture the innate performance of each pretrained model without confounding it with additional parameters or training.

\subsection{Text Encoders}

We experimented with two different multilingual text models, the Fairly Multilingual ModernBERT Embedding Model -- Belgian Edition (FMML-BE) ~\cite{remy-2025-fmmb-be} and Llama 3.2 with 1B parameters~\cite{grattafiori2024llama3herdmodels}. These are strong multilingual text-based models for transcript representation. As we did with the speech models, we froze the models and added minimal aggregation parameters on top, to try and capture the effect of the pretraining process rather than fine-tuning.
 
\section{Experiments}

Here we detail the experiments we carried out on Parkinson's disease detection using a variety of models and settings.

\subsection{Dataset}

For our experiments with Parkinson's disease, we used a set of speech recordings from the Quebec Parkinson Network (QPN)~\cite{gan2020quebec} --- 208 patients and 52 controls. The demographic breakdown of the study subjects can be seen in Table~\ref{tab:qpn_breakdown}. A portion of these speech data have previously been used to show that articulation impairments in patients with PD are associated with aberrant activity in the left inferior frontal cortex~\cite{wiesman2023aberrant}. All patients and controls were recorded in the same way with a headset microphone near the mouth and in a quiet room. The majority of patients were recorded in the ON medication state, meaning they had taken their prescribed dopaminergic treatment (e.g., levodopa) prior to the session. Prior work has shown that dopaminergic medication has limited effects on speech production in Parkinson’s disease~\cite{cavallieri2021dopaminergic}.

The subjects were asked to perform a number of tasks, such as sustained phonation, reading a passage, recalling a memory, and describing a picture (DPT). The latter was selected as initial experiments showed the best discrimination performance --- likely due to the cognitive demands of the task~\cite{bocanegra2015syntax}.

From these data, we selected 32 subjects to form a demographically balanced test set, which contains equal numbers of patients and controls, native French and English speakers, and men and women. To ensure that patients were in the early stages of the disease, we selected subjects with an average time between diagnosis and speech recording of 3.1 years, compared to 4.1 years in the training set. Finally, we primarily selected patients with mild motor symptoms, as indicated by Unified Parkinson's Disease Rating Scale (UPDRS) II scores of 12 or less and UPDRS III scores of 32 or less ~\cite{martin2015parkinson's}.
For the training set, the data are not balanced. We use sampling and weighting strategies to address the patient status and sex imbalances. 

As for language, although native French speakers outnumber other speakers by a ratio of 2.2 to 1, 59\% of them also speak English and for these subjects we have recordings of tasks in both languages. This makes the final language distribution more balanced. 

\begin{table}
    \centering

    \begin{tabular}{ll}
        \toprule
        \multicolumn{2}{c}{\textbf{208 Patients}} \\
        \midrule
        Sex & 133 Males -- 75 Females \\
        Age & Mean 65.9 yr. -- Stddev 8.7 yr. \\
        First language & French 73\% -- English 13\% -- Other 14\% \\
        UPDRS III & Mean 29.6 -- Stddev 13.6 \\ 
        \midrule
        \multicolumn{2}{c}{\textbf{52 Controls}} \\
        \midrule
        Sex & 16 Males -- 36 Females \\
        Age & Mean 63.7 yr. -- Stddev 8.8 yr. \\
        First language & French 44\% -- English 34\% -- Other 22\% \\
        \bottomrule
    \end{tabular}
    \caption{Demographic breakdown of the QPN speech dataset}
    \label{tab:qpn_breakdown}
\end{table}

\subsection{Speech Experimental Setup}

We experimented with a minimal configuration of parameters on top of each frozen encoder in order to represent the effects of pretraining as much as possible. For each encoder, we train a small classification module consisting of the following layers: linear, attention pooling across time, linear, output (binary). The linear layers have either 768 or 1280 neurons, which are the output sizes of our smallest and largest models. We found from experimentation that the ASR models perform better with the 768-dim classification and the SSL models perform better with the 1280-dim classification, which is the setting we used for our experiments. Between linear layers, we added dropout at a rate of 0.2 and leaky ReLU activations.

We trained all models with an Adam 8-bit optimizer and used a learning rate of 0.0001 with linear warm-up for 2 epochs and cosine cool-down for the remaining 18 epochs. Every epoch consists of 1024 samples with 32 samples per batch.

We used sampling and weighting techniques to address the data imbalances in terms of sex and subject type. For all experiments, we equally sample from the four categories created by the intersection of these two conditions. In addition to adjusting the sampling for sex and patient status, we also re-weighted samples, at 0.7 for majority types and 1.5 for minority types. We sample 1024 utterance chunks of 30 seconds or less each epoch.

To reduce the incidence of overfitting on the over-sampled data, we introduced data augmentations on the waveforms of 90\% of batches. We combined two augmentations: adding background noises at random signal-to-noise ratios between 0 and 15 and dropping between 2 and 5 random frequencies via notch filters. For the sustained phonation task we additionally zeroed out random segments. The augmentations are dynamically applied~\cite{speechbrain}, and can only obscure important features on a fraction of the data presentations, as a regularization.


In addition to pretrained models, we provide results on mel filterbank and extracted vocal features, which serve as baselines for comparing the pretrained model features. The vocal features are a combination of autocorrelation-based features (f0, harmonicity, glottal-to-noise-excitation ratio~\cite{godinollorente2010gne}), period-based features (jitter, shimmer), spectral-based features (e.g. skew, flux), and the first 4 mel-frequency cepstral coefficients.

\subsection{Transcript Experimental Setup}

To generate the transcripts, we passed the audio through Google Cloud's Chirp2 model~\cite{zhang2023google}, which supports Canadian French, and verified that the systems appropriately created transcripts. While Google's model already infers short pauses using commas and periods, we also insert ellipses ("...") to any gaps in the transcript that exceed two seconds, following the method in~\cite{yuan2021pauses}.

For the QPN dataset, we had manually transcribed 101 of the picture description recordings. This allowed us to estimate the Word Error Rate (WER) of Google's Chirp2 model at 21.1. We further broke down the WER by language and found a WER of 25.7 for French transcripts compared to just 9.7 for English transcripts. This is likely due to a larger amount of training data available for English ASR ~\cite{zhang2023google}. Finally, we also separated the WER by subject type and found that the ASR model made more errors when transcribing PD patients with a WER of 23.2 compared to a WER of 14.5 for healthy controls.

On top of each frozen text encoder we train a minimal classification model of size 768 using the same methodology as the audio encoders. In the input, we dynamically augment the transcripts by dropping 10\% of words, swapping 5\% of the words, and replacing 20\% of the transcripts with their translation from Google Translate.

\section{Results}

\begin{table}[t!]
    \centering
    \begin{tabular}{lcccccccc}
        \toprule
         & \textbf{SVP} & \textbf{DPT} & \textbf{DPT} & \textbf{DPT} \\
        \textbf{Encoder} & \textbf{F1} &\textbf{Fr F1} & \textbf{En F1} & \textbf{F1} \\
        \midrule
        SB VocalFeats        & 59.2 & 74.7 & 59.0 & 65.9 \\
        Filterbank n=80      & 50.5 & 66.2 & 54.0 & 60.0 \\
        OS eGeMAPSv02        & 50.5 & 52.6 & 54.6 & 53.7 \\
        \midrule
        FMMB-BE              & - & 75.5 & 66.0 & 70.5 \\
        Llama 3.2 1B         & - & 70.8 & 57.7 & 64.7 \\
        \midrule
        Whisper Small        & 55.5 & 86.4 & \textbf{75.3} & \textbf{80.8} \\
        Whisper Medium       & 51.0 & 82.2 & \textbf{78.4} & \textbf{80.2} \\
        Whisper Large v3     & 52.6 & 80.2 & 62.5 & 72.1 \\
        Whisper Small.en     & 49.5 & 73.5 & 65.9 & 70.1 \\
        Crisper Whisper      & 52.9 & 75.3 & 61.1 & 68.5 \\
        Parakeet CTC 0.6B    & 49.4 & 72.7 & 52.8 & 64.8 \\
        \midrule
        XEUS                 & 60.6 & 83.9 & 69.3 & 76.5 \\
        WavLM Base+          & 52.6 & 81.2 & 68.5 & 75.2 \\
        WavLM Large          & 49.0 & 78.9 & 70.7 & 75.1 \\
        HuBERT Base          & 45.8 & 80.0 & 66.6 & 71.6 \\
        HuBERT Large         & 48.4 & 77.8 & 34.7 & 58.5 \\
        wav2vec 2 Base       & 51.0 & 78.7 & 56.4 & 68.8 \\
        \midrule
        HuBERT Base - AS     & \textbf{67.5} & 78.1 & 52.5 & 67.2 \\
        wav2vec 2 Base - AS  & 57.1 & 64.6 & 56.1 & 60.2 \\
        \bottomrule
    \end{tabular}
    \caption{Parkinson's disease detection accuracy scores using spoken picture descriptions. Listed tasks are Sustained Vowel Phonation (SVP) and Picture Description Task (DPT). The latter was conducted in both French (Fr) and English (En). Listed sections are divided by encoder type: (1) Interpretable features, (2) Transcript models, (3) ASR models, (4) SSL models (speech), and (5) SSL models (AudioSet). Results are bolded where a one-sided t-test shows that the top system(s) achieve significantly better accuracies than the next-best system at $p<0.5$.}
    \label{tab:speech_for_pd}
\end{table}

The results of our experiments can be seen in Table~\ref{tab:speech_for_pd}, our insights from this table are listed below:

First, the picture description task is more informative for the detection task than the sustained vowel phonations. We note that for most systems, the performance on the sustained vowel phonation task is near chance (50\%), with the exception of our best-performing system in each category, as well as the AudioSet-trained models.

Second, the best text-based model performance (70\%) is similar to the best SVP performance (68\%) as well as the best vocal features on the same data (66\%). From this, we conclude that language features are approximately as discriminative as vocal features. One note here is that we used manual transcripts where available and ASR otherwise. When using only ASR transcripts, the high rate of errors for French led to reduced performance, around 60\% F1 score for FMMB-BE. However, this doesn't affect the main point -- words and punctuation can be an important source of information for the early detection of PD.

Third, Whisper performed the best out of all tested models. We note that the English-only variant of Whisper performed much worse than the multilingual version. This suggests that something about the multilingual training process provides a significant boost to detection accuracy. This could be due to one of the additional tasks (language classification or translation), or the added multilingual data could provide feature robustness in the presence of variations in speech production.

Fourth, for SSL models, the difference between monolingual and multilingual trained encoder performance on the picture description task is much smaller than for Whisper. This could indicate that these models pay less attention to linguistic cues and more attention to voice changes, such as articulation. This point is reinforced by the stronger performance of XEUS on the phonation task. These results could potentially indicate a greater robustness to language variations, although this might be offset by reduced robustness to noise~\cite{la2024exploiting}.

Finally, for the sustained vowel phonation task, SSL models pretrained on Audioset perform much better than SSL models trained on speech. This suggests that that since relatively little language information is available from the SVP task, reasonable performance requires a more domain-appropriate approach.

\section{Conclusion}

Our experiments provide evidence that language is as informative as voice measures for the early detection of Parkinson's disease, although neither alone is sufficient for the best detection. We observe strong performance from the multilingual Whisper encoder but see relatively poorer performance for the English-only variant, suggesting that exposure to multilingual data and objectives are a crucial part of the pretraining process. By contrast, self-supervised models showed less language dependence and stronger performance on the SVP task, likely indicating a greater dependence on vocal features.

In conclusion, this paper finds that future work to develop biomarkers for Parkinson's disease from speech would do well to prioritize tasks that use the language areas of the brain, as well as to consider the linguistic aspects of model pretraining when selecting foundation models.

\bibliographystyle{IEEEtran}
\bibliography{refs}

\end{document}